\documentclass[submission, Phys]{SciPost}
\hypersetup{
    colorlinks,
    linkcolor={black!50!black},
    citecolor={blue!50!black},
    urlcolor={blue!80!black}
}
\usepackage{physics}
\usepackage{graphicx}
\usepackage{amsmath,amscd,amssymb,amsthm,bbm,hyperref,tikz,float,bbold}  
\graphicspath{{images/}}
\usepackage{subcaption}

\usepackage[bitstream-charter]{mathdesign}
\DeclareSymbolFont{usualmathcal}{OMS}{cmsy}{m}{n}
\DeclareSymbolFontAlphabet{\mathcal}{usualmathcal}

\begin{document}

\begin{center}{\Large \textbf{
Towards a phase diagram of the topologically frustrated XY chain\\
}}\end{center}

\begin{center}
Daniel Sacco Shaikh\textsuperscript{1$\star$},
Alberto Giuseppe Catalano\textsuperscript{2,3},
Fabio Cavaliere\textsuperscript{1,4},
Fabio Franchini\textsuperscript{3 $\dagger$},
Maura Sassetti\textsuperscript{1,4} and
Niccolò Traverso Ziani\textsuperscript{1,4}
\end{center}

\begin{center}
{\bf 1} Dipartimento di Fisica, Università degli Studi di Genova, via Dodecaneso 33, 16146, Genova, Italy.
\\
{\bf 2} Division of Theoretical Physics, Ruđer Bošković Institute, Bijenička cesta 54, 10000 Zagreb, Croatia.
\\
{\bf 3} Université de Strasbourg, 4 Rue Blaise Pascal, 67081 Strasbourg, France.
\\
{\bf 4} CNR SPIN, via Dodecaneso 33, 16146, Genova, Italy.
\\
${}^\star$ {\small \sf s4526348@studenti.unige.it}
$\qquad{}^\dagger$ {\small \sf fabio.franchini@irb.hr}
\end{center}

\begin{center}
\today
\end{center}
\section*{Abstract}
{\bf
Landau theory's implicit assumption that microscopic details cannot affect the system's phases has been challenged only recently in systems such as antiferromagnetic quantum spin chains with periodic boundary conditions, where  topological frustration can be induced. In this work, we show that the latter modifies the zero temperature phase diagram of the XY chain in \textcolor{black}{ a transverse magnetic} field by inducing new quantum phase transitions. In doing so, we come across the first case of second order boundary quantum phase transition characterized by a quartic dispersion relation. Our analytical results are supported by numerical investigations and lay the foundation for understanding the phase diagram of this frustrated model.}\\
\noindent\rule{\textwidth}{1pt}
\tableofcontents\thispagestyle{fancy}
\noindent\rule{\textwidth}{1pt}
\vspace{10pt}

\section{Introduction}
\label{sec:intro}
According to Landau theory \cite{Landau_Phase_Transitions,LANDAU1980446}, one of the milestones of classical statistical mechanics, phases separated by a phase transition can be distinguished by a change in the behavior of a local order parameter, enabling to assess the macroscopic order in a system's phase. In particular, a non-zero local order parameter is a manifestation of spontaneous symmetry breaking \cite{GoldstoneWeinbergSalam,Anderson_SSB, Notes_SSB} and the establishment of a  macroscopic order that explicitly violates one of the specific symmetries of the theory.  Symmetry breaking happens when crossing a critical point, where the system can reorganize itself modifying its macroscopic behavior \cite{Goldenfeld}. Because of its success, Landau theory has firstly been borrowed without changes when dealing with quantum phases of matter \cite{sachdev_2011}. However, it soon became clear that the complexity of quantum many-body systems was not fully captured by this theory. For example, it cannot explain nematic phases \cite{NicShannon,GiampaoloNematic} (where the breaking of a symmetry is not univocally associated to a single order parameter) and topological phases \cite{fradkin_2013, Witten_TopologicalPhases} (described by non-local order parameters). This  motivated calls for an extension of the Landau theory to include the description of those quantum phenomena which do not have any classical counterpart.\\

\noindent Also a second assumption of the Landau theory could probably be challenged: its implicit assumption that {\em microscopic} changes in the system are {\em negligible} in determining its {\em thermodynamic properties}, hence the prescription to take the thermodynamic limit before doing any calculation. As a consequence, any change in the boundary conditions is considered to be irrelevant when dealing with the bulk, macroscopic behavior. Recently, however, it has been proven that at a quantum level this assumption misses part of the physics, precisely in the case of antiferromagnetic quantum spin chains with discrete symmetries and \textit{frustrated boundary conditions} (FBC) \cite{Giampaolo_2019}, namely periodic boundary conditions with an odd number of sites. Such a peculiar choice of boundary conditions implies that the simultaneous minimization of all local interactions in the Hamiltonian is not compatible with the system's geometry, giving rise to geometrical frustration \cite{J_Vannimenus_1977, Chalker2011, Moessner_Ramirez}. In the following, this kind of geometrical frustration induced by the lattice's topology \cite{QPT_induced_by_TF} will be called \textit{topological frustration} (TF), a name which has also been more rigorously justified by Torre \textit{et al.} in \cite{torre2023longrange}.\\
Over the years, extensively frustrated systems have been studied at large and it has been shown that they exhibit peculiar and exotic physical behaviors \cite{Moessner_Ramirez,Moessner_GeometricalFrustration,lacroix}  which are very different with respect to those of their non-frustrated counterparts, both at a classical and at a quantum level. One of the main features of classical frustrated systems is the presence of highly degenerate ground state (GS) manifolds in the limit of a large number of sites \cite{Moessner_GeometricalFrustration}. This property can also be found in the simple case of a classical Ising chain with $N$ sites and FBC, which can be thought as a single building block of an extensively frustrated system \cite{Giampaolo_2019}. The GS space in the frustrated case is $2N$ times degenerate and spanned by kink states, with a single pair of nearest neighbor spins ferromagnetically aligned while the other $N - 1$ pairs are antiferromagnetically aligned. In contrast, without frustration the GS space would have degeneracy 2 and would be spanned the two Neél states (perfectly staggered antiferromagnetic states). At a quantum level, the extensive degeneracy introduced by FBC is lifted: the GS, at a perturbative level, is given by a superposition of kink states and becomes part of a Galilean band of gapless excitations \cite{Dong_2016,Dong_XY_17} in a phase that would have been gapped without frustration. Intuitively, one could still be led to think that the contributions of the boundary terms become irrelevant for sufficiently large systems. However, such an expectation has already been challenged several times \cite{Cabrera86, Cabrera1987,Barber87, Campostrini_Pellissetto_Vicari,Campostrini_2015_Jstat, Dong_2016, Dong_XY_17, Dong2018}. For instance, Campostrini \textit{et al.} \cite{Campostrini_Pellissetto_Vicari} showed that by tuning the strength of {\em a single bond defect} in an odd length ferromagnetic quantum Ising ring in a transverse magnetic field the system can be driven across a quantum phase transition (QPT) which separates a gapped magnetic phase from a gapless (but non relativistic) kink phase. Relevantly, at the  transition point this model can be exactly mapped to the quantum Ising chain with FBC \cite{Giampaolo_2019}. In the last few years important progresses have been made along this research line, particularly studying topologically frustrated quantum spin chains \cite{Torre_loschmidt,QPT_induced_by_TF,fate_of_local_order,Frustration_of_being_odd,Franchini_EffectsOFDefects,Maric_topological,TF_modify_the_nature_of_QPT,Catalano22,Odavi__2023,torre2023longrange}, and the results indicate that FBC can completely change the zero temperature phase diagram and the low-energy properties. In particular, Mari\'c \textit{et al.} \cite{TF_modify_the_nature_of_QPT} provided an example of a quantum spin chain with FBC showing a complete destruction of the local order at both sides of a QPT, proving that TF can modify the nature of a critical point. The same authors have also shown \cite{fate_of_local_order} that the disappearance of the local order is a common fact when the wider class of translational invariant topologically frustrated Ising-type spin chains endowed with the parity symmetries in all three spin directions are considered. Furthermore, two examples of boundary quantum phase transition (b-QPT) induced by TF have been put forward \cite{QPT_induced_by_TF, articolo_tesi} and the role of FBC on the modification of the local order at both sides of the new b-QPT has been analyzed \cite{QPT_induced_by_TF,Frustration_of_being_odd}. It has also been shown \cite{Franchini_EffectsOFDefects} that phases characterized by a non vanishing magnetization in the thermodynamic limit are resilient to the introduction of an antiferromagnetic defect (spoiling translational invariance), paving the way for possible experimental observations.\\

\noindent The aim of this work is to investigate how TF modifies the zero temperature phase diagram of the XY quantum spin chain in a transverse magnetic field. After a detailed analysis on the drastic changes of the GS properties due to the imposition of FBC, we will analytically prove that TF induces new b-QPTs: an additional first order b-QPT appears in correspondence of the conformal lines of the unfrustrated case, a first order b-QPT with Galilean dispersion relation shows up at zero field and, finally, a second order b-QPT with a quartic dispersion relation is induced in a region of the phase diagram that would have been otherwise ordered and gapped without frustration. To the best of our knowledge, the latter phase transition, which separates two gapless regions of the parameter space with strikingly different GS behavior, is the first case of second order QPT characterized by a dispersion relation which is neither relativistic nor Galilean, but quartic. We also report numerical calculations that support the validity of our analytical computations.\\

\noindent The paper is organized as follows: in Section \ref{sec::model} we introduce the model and the essential details about its exact diagonalization. In Section \ref{sec::Signatures of topological frustration} the general features of its GS, focusing on the novelties induced by FBC, are discussed. In Sections \ref{sec:firstorder} and \ref{sec:secondorder} we  present and comment our results about b-QPTs induced by FBC. In Section \ref{sec::conclusion} we draw our conclusions.

\section{Model}\label{sec::model}
The model we study is the XY chain in a transverse magnetic field \cite{katsura,McCoyXY1,McCoyXY2,McCoyXY3}, which is defined by the following Hamiltonian
\begin{align}
    H
    =&\frac{J}{2}\sum_{j=1}^N \left[\left(\frac{1+\gamma}{2}\right)\sigma_j^x\sigma_{j+1}^x + \left(\frac{1-\gamma}{2}\right)\sigma_j^y\sigma_{j+1}^y + h\sigma_j^z\right]\label{xy model},
\end{align}
where $N$ is the number of sites, $J$ is the energy scale, $\gamma$ is a parameter quantifying the anisotropy of nearest neighbour spins interactions, $h$ is an external magnetic field along $z$ direction and $\sigma_j^\alpha$ (with $\alpha=x,y,z$) are the Pauli operators
on the $j$-th lattice site satisfying
\begin{equation}
    [\sigma_j^\alpha, \sigma_k^\beta] = 2i \epsilon^{\alpha\beta\gamma} \sigma_j^\gamma \delta_{jk}.
\end{equation}
In this paper we will focus on the antiferromagnetic case and thus choose $J>0$. In particular, here and in the rest of the paper we will set $J=1$.\\

\noindent The FBC are imposed enforcing periodic boundary conditions with an {\em odd} number of lattice sites:
\begin{equation}\label{PBC}
    \sigma_{j+N}^\alpha \equiv \sigma_j^\alpha \qquad \alpha=x,y,z\, .
\end{equation}
Without loss of generality, we will restrict our analysis only to $h,\gamma \geq 0$ owing to the symmetries of the XY model. Note that when $\gamma=1$ the above model reduces to the frustrated quantum Ising chain, which has been studied in \cite{Dong_2016, articolo_tesi, Torre_loschmidt}.\\

\noindent The exact diagonalization of the Eq. \eqref{xy model} is well known \cite{LIEB1961407,katsura,NIEMEIJER1967377,NIEMEIJER1968313} but since imposing FBC requires some care. Therefore, the explicit solution is reported in App. \ref{app:diagonalization}, where we also compare the frustrated and unfrustrated regimes. Here, we only report the key element necessary to state our results, {\em i.e.} the parity operators
\begin{equation}\label{parity operators}
    \Pi^\alpha \equiv \bigotimes_{l=1}^N \sigma_l^\alpha \qquad\alpha=x,y,z.
\end{equation}
The system Hamiltonian is invariant under the action of the $z$-parity operator $[H,\Pi^z] =0$, so that the decomposition
\begin{equation}
    H = \frac{1+\Pi^z}{2}H^+ \frac{1+\Pi^z}{2} + \frac{1-\Pi^z}{2}H^- \frac{1-\Pi^z}{2}
    \label{parity constraints}
\end{equation}
holds. The explicit expressions for $H^{\pm}$ (see App. \ref{app:diagonalization}) are
\begin{align}
    &H^+ = \sum_{q\in \Gamma^+\setminus\{\pi\}}\epsilon(q)\left(\chi_q^{\dagger} \chi_q-\frac{1}{2}\right)-\epsilon(\pi)\left(\chi_\pi^{\dagger} \chi_\pi-\frac{1}{2}\right) \label{H^+} \qquad\mbox{if}\qquad h\geq 0,\\
    \nonumber\\
    &H^- = \begin{cases}
    \sum_{q\in \Gamma^-}\epsilon(q)\left(\chi_q^\dagger \chi_q-\frac{1}{2}\right) &\quad\mbox{if}\quad  0\leq h< 1\\[1 ex]
    \sum_{q\in \Gamma^-\setminus\{0\}}\epsilon(q)\left(\chi_q^\dagger \chi_q-\frac{1}{2}\right) -\epsilon(0)\left(\chi_0^\dagger \chi_0-\frac{1}{2}\right) &\quad\mbox{if}\quad h\geq 1,
    \label{H^-}
    \end{cases}
\end{align}
where $\chi_q$ are Bogoliubov fermionic operators and
\begin{align}
    &\Gamma^- = \left\{\frac{2\pi}{N}k\right\}, \qquad \Gamma^+ = \left\{\frac{2\pi}{N}\left(k+\frac{1}{2}\right)\right\} \qquad k=0,...,N-1\,,
    \label{Gamma sets}\\
    \nonumber\\
    &\epsilon(q)= \sqrt{(h-\cos q)^2 + \gamma^2 \sin^2q}.\label{epsilon}
\end{align}
The study of the sets \eqref{Gamma sets} and the spectrum \eqref{epsilon} plays a crucial role in the derivation of even-odd effects in this model.
\section{Signatures of topological frustration on the GS}\label{sec::Signatures of topological frustration}
We now comment the general features of the frustrated XY chain, starting from its GS. We call $\ket{GS'^\pm}$, $\ket{GS^\pm}$  and $\ket{GS}$ the most general elements of the GS of $H^\pm$, $\frac{1\pm\Pi^z}{2}H^\pm$ and $H$ respectively. An analogous notation for the corresponding energies will also be employed. Furthermore, $\ket{0^\pm}$ denotes the vacuum of $\chi_q$. The strategy to obtain the GS is the following: after having identified $\ket{GS'^\pm}$, the states $\ket{GS^\pm}$ can be found applying $\frac{1\pm\Pi^{z}}{2}$. This in turn allows to find the GS energy $E=\min\{E^+,E^-\}$ and the corresponding set of $\ket{GS}$. In the thermodynamic limit $N\to\infty$ this procedure is fully analytical, while at finite $N$ one has to resort to numerical methods.\\
\subsection{In the absence of frustration}
As shown in App. \ref{app:diagonalization}, in the absence of FBC the system is completely equivalent to its ferromagnetic counterpart \cite{Franchini_2017, depasquale}\textcolor{black}{,} then $\ket{GS^+}=\ket{GS'^+}=\ket{0^+}$ and $\ket{GS^-}=\ket{GS'^-} = \chi^\dagger_\pi\ket{0^-}$ alternate as GSs and with corresponding energies 
\begin{align}
    &E^+=  -\frac{1}{2}\sum_{q\in\Gamma^+}\epsilon(q),\label{E^+ even}\\
    \nonumber\\
    &E^- = \begin{cases} -\frac{1}{2}\sum_{q\in\Gamma^-}\epsilon(q) + \epsilon(0) &\quad\mbox{if}\quad h\geq 1\\
    -\frac{1}{2}\sum_{q\in\Gamma^-}\epsilon(q) &\quad\mbox{if}\quad 0\leq h<1.\label{E^- even}
    \end{cases}
\end{align}
Furthermore one can observe that there are $\lfloor N/2\rfloor$ level crossings in the first quarter of the parameter space and that $h^2+\gamma^2=1$ is an exact doubly degeneracy line for all sizes \cite{depasquale}. To determine precisely the GS at fixed $(h,\gamma)$ one has to compare the energies \eqref{E^+ even} and \eqref{E^- even}. It can be shown that at large $N$ the energy gap between these two states closes exponentially \cite{Cabrera1987,Damski_2014}, giving rise to a doubly degenerate manifold which spontaneously breaks $\mathbb{Z}_2$ symmetry \cite{Franchini_2017}.
\subsection{The frustrated case}
The presence of TF modifies significantly the scenario, particularly for $\abs{h}<1$, where one cannot chose $\ket{GS'^+}$ and $\ket{GS'^-}$ as ground states, with their $z$-parity being equal to $-1$ and $+1$ respectively. In other words, the lowest energy states of $H^+$ and $H^-$ are not compatible with the $z$-parity constraint \eqref{parity constraints}. As a consequence, the GS with FBC can be interpreted as a single excitation in the system with respect to a fermionic vacuum state. This fact can be seen from Eqs. \eqref{H^+}, \eqref{H^-}, \eqref{epsilon} imposing the parity constraint \eqref{parity constraints}. 
\begin{figure}[h!]
    \centering
    \includegraphics[scale=1]{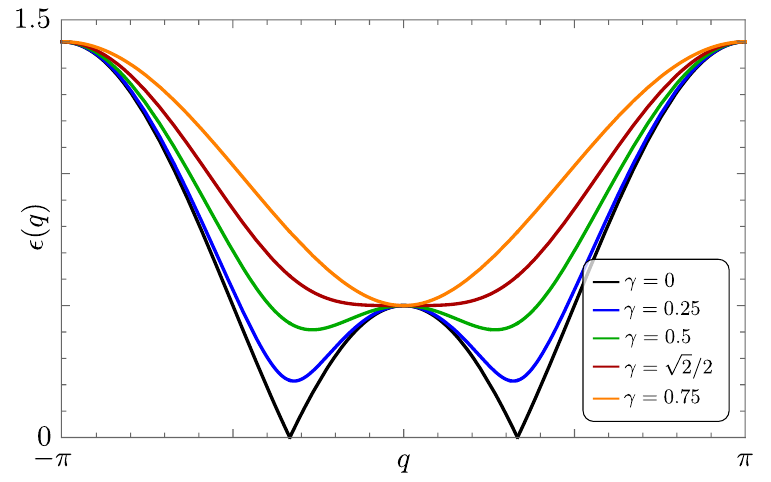}
    \caption{Plot of the dispersion relation $\epsilon(q)$ as a function of $q\in [-\pi,\pi]$ for $h=0.5$ and different values of $\gamma$: $\gamma= 0$ (black), $\gamma=0.25$ (blue), $\gamma=0.5$ (green), $\gamma=\gamma^* (0.5)=\sqrt{2}/2$ (red) and $\gamma=0.9$ (orange).}
    \label{fig:grafico_epsilon_h<1}
\end{figure}
From the shape of $\epsilon(q)$ when $0<h<1$ -- see Fig. \ref{fig:grafico_epsilon_h<1} -- we can divide this region of the parameter space in two subregions, separated by a line $\gamma^*(N,h)$: one occurring when $0<\gamma< \gamma^*(N,h)$ in which the set \textcolor{black}{$\{\epsilon(q)\}_{q\in\Gamma^\pm}$} has \textcolor{black}{two symmetric absolute} minima and one for $\gamma>\gamma^*(N,h)$ where it has the only one minimum \textcolor{black}{$\epsilon(0)$}. It can be shown that \cite{Dong_XY_17}
\begin{equation}
\lim_{N\to\infty}\gamma^*(N,h)\equiv\gamma^*(h) = \sqrt{1-h}.    
\end{equation}

\noindent In the subregion $0<h<1,\,\gamma> \gamma^*(h)$ the dispersion relation $\epsilon(q)$ always has the absolute minimum at $q=0$ (as in Fig. \ref{fig:grafico_epsilon_h<1}), hence the GS in the thermodynamic limit is not degenerate and given by $\chi_0^\dagger\ket{0^-}$ with energy
\begin{equation}
    E  \simeq E_0+1-h\,,
\end{equation}
where
\begin{equation}
    \label{eq::E0}
    E_0\textcolor{black}{\equiv}-\frac{N}{4\pi}\int_0^{2\pi}\epsilon(q)\dd q
\end{equation}
is the GS energy in the absence of frustration \cite{Franchini_2017}. Moreover, the GS is now part of a band of states and the system is gapless, with the energy gap closing as $1/N^2$.\\

\noindent In the subregion $0<h<1,\,0\leq\gamma<\gamma^*(h)$ the dispersion relation has always two symmetric absolute minima (as shown in Fig. \ref{fig:grafico_epsilon_h<1}) at $q=\pm p(h,\gamma)$ with \cite{Dong_XY_17}
\begin{equation}
    p(h,\gamma) = \arccos\left({\frac{h}{1-\gamma^2}}\right),
\end{equation}
hence the GS depends on the exact position in the parameter space $(h,\gamma)$ and is given by
\begin{equation}
    GS^+ = \mbox{span}\bigg\{\chi^\dagger_{p_+(h,\gamma)}\chi_\pi^\dagger\ket{0^+},\, \chi^\dagger_{-p_+(h,\gamma)}\chi_\pi^\dagger\ket{0^+}\bigg\},
\end{equation}
or
\begin{equation}
    GS^- = \mbox{span}\bigg\{\chi^\dagger_{p_-(h,\gamma)}\ket{0^-},\, \chi^\dagger_{-p_-(h,\gamma)}\ket{0^-}\bigg\}
\end{equation}
where we denoted with $p_+(h,\gamma)$ and $p_-(h,\gamma)$ respectively the closest elements to $p(h,\gamma)$ in $\Gamma^+$ and $\Gamma^-$. It is interesting that, as already pointed out in \cite{Catalano22}, in the thermodynamic limit the GS fidelity does not vanish only if we move along one of the parabolas $h=c(1-\gamma^2)$ with $c\in [0,1]$, where the minima of $\epsilon(q)$ remain fixed. Hence the frustrated system exhibits a behavior similar to a model with a continuous symmetry (e.g. the XXZ chain in its paramagnetic phase \cite{Franchini_2017}). 
\\In the thermodynamic limit, the energy of the GS in this particular region of the parameter space is
\begin{equation}
    E \simeq E_0+\gamma\sqrt{\frac{h^2+\gamma^2-1}{\gamma^2-1}}\,.
\end{equation}
Moreover, the energy gap separating the finite size two-fold degenerate ground state manifold with the first excited one closes exponentially with $N$ in the thermodynamic limit \cite{Catalano22}, giving rise to a gapless GS with degeneracy 4, spanned by states with opposite $z$-parity. Note that the cardinality of the two sets $\Gamma^+$ and $\Gamma^-$ scales proportionally to the number of sites. As a consequence, increasing $N$ the number of level crossings between twice degenerate manifolds with opposite $z$-parity and non-vanishing momenta \cite{Catalano22} increases. In particular, observe that the mirror symmetry of our theory, \textit{i.e.} the invariance of the Hamiltonian in Eq. \eqref{xy model} under the transformation mapping the spin operator $\sigma_j^\alpha$ into $\sigma_{2k-j}^\alpha$ where $k$ is a generic site of the chain, implies that if the GS momentum is not zero, then the ground state manifold is a two-fold degenerate manifold spanned by single particle excitations with equal and opposite momenta, as proven in Ref. \cite{QPT_induced_by_TF}.\\

 \noindent Finally, for $h\geq 1$, the GS energy in the thermodynamic limit is simply $E=E_0$. 

\noindent Some comment on the line $h=0$ of the parameter space, which has already been studied in detail \cite{QPT_induced_by_TF, Frustration_of_being_odd,fate_of_local_order}, are in order. Note that here we have $[H,\Pi^\alpha] =0$ (\textit{i.e.} each parity operator is a symmetry), and that the parity operators satisfy the non-commuting algebra $\{\Pi^\alpha, \Pi^\beta\} = 2\delta_{\alpha,\beta}$ in the odd $N$ case. As a consequence, for a state $\ket{\psi}$ with definite energy and $z$-parity,  $\Pi^x\ket{\psi}$ has the same energy but opposite $z$-parity. Hence, for $h=0$ and odd $N$, even at finite size each energy level is always degenerate an even number of times (Kramers degeneracy \cite{Tasaki_book}) and spontaneous symmetry breaking occurs even without taking the thermodynamic limit. Taking advantage of this extra symmetry, the GS magnetization (\textit{i.e.} the expectation value of $\sigma_j^\alpha$) and the two-spin correlation functions have been computed analytically \cite{QPT_induced_by_TF,Frustration_of_being_odd}. For $\gamma>1$, where the ground state manifold is gapless and twofold degenerate, the average magnetization is constant and is suppressed by the total system size as $1/N$, a phenomenon called \textit{mesoscopic ferromagnetic magnetization} \cite{Frustration_of_being_odd}. On the other hand, for $\gamma<1$ the GS is still gapless but four times degenerate and, as previously observed, the operators $H$, $\Pi^z$ and the translation operator $T$ (defined as $T^\dagger \sigma_j^\alpha T \equiv \sigma_{j+1}^\alpha$) form a complete set of compatible operators. As a consequence, there are two possibilities: choosing a translationally invariant state and obtaining the mesoscopic ferromagnetic magnetization or giving up the translational invariance obtaining a state whose magnetization looks like the staggered one but changes incommensurably over the chain \cite{QPT_induced_by_TF}. Contrary to the former, the latter incommensurate configuration has been proven to be resilient even in the presence of antiferromagnetic-type defects breaking the translational symmetry of the model \cite{Franchini_EffectsOFDefects}. Moreover, the system undergoes a first order b-QPT at the point $\gamma=1$ \cite{QPT_induced_by_TF}. Such a QPT is a consequence of the FBC and would be absent for other boundary conditions. The results about the frustrated XY chain \textcolor{black}{at} $h=0$ have been generalized \cite{fate_of_local_order} to the wider class of translationally invariant weakly-frustrated spin-$1/2$ Hamiltonians with a dominant antiferromagnetic Ising-type interaction in one direction and commuting with all three parity operators \eqref{parity operators}. In this case the expectation values of all local operators (with support on a finite fixed number of sites) have been shown to decay to zero at least algebraically with the system size unless the GS manifold contains at least two states whose momenta differ by $\pi$ in the thermodynamic limit. Due to the previously discussed symmetries, this can be possible only for GS manifolds which are at least four times degenerate. Note that local order implies breaking the translational symmetry: this is the case of the XY chain at zero field with $\gamma<1$ \cite{QPT_induced_by_TF}.
\section{First order b-QPTs}
\label{sec:firstorder}
In this Section we study the possible presence of curves in the $(h,\gamma)$ plane at which the system undergoes first order b-QPTs, which are non-extensive discontinuities in the first derivative of the GS energy \cite{MatthiasVojta_2003}. In the previous Section we computed the GS energy in the large $N$ limit to be of the form $E=E_0+\Delta E$ with $E_0$ defined in Eq. \eqref{eq::E0} and
\begin{align}
    \Delta E(h,\gamma) = \begin{cases}
    0  &\mbox{if}\quad \abs{h}\geq 1\\[1.5 ex]
    \abs{\gamma}\sqrt{\frac{h^2+\gamma^2-1}{\gamma^2-1}} &\mbox{if}\quad \abs{h}< 1, \,\abs{\gamma} < \gamma^*(\abs{h})\\[1.5 ex]
    1-\abs{h}  &\mbox{if}\quad \abs{h}< 1, \,\abs{\gamma} \geq \gamma^*(\abs{h})
    \end{cases}\label{def DeltaE}
\end{align}
This last quantity represents the energy difference induced by TF. Note that the correction $\Delta E$, which is a distinguishing feature of FBC, does not scale with the number of sites. As a consequence, any possible \textcolor{black}{new} discontinuity of $E$ cannot be extensive.\\
From Eq. \eqref{def DeltaE} we observe that the system undergoes a first order b-QPT at $h=1$ with $\gamma>\textcolor{black}{0}$ and at $h=0$ with $\gamma\geq 1$. Moreover, when $0\leq h\leq 1$
\begin{equation}\label{first order b-QPT gamma=0}
    \lim_{\gamma\to0^+}\pdv{E}{\gamma}-\lim_{\gamma\to0^-}\pdv{E}{\gamma} 
    = 2\sqrt{1-h^2},
\end{equation}
so there is another first order b-QPT at $\gamma=0$ with $0\leq h< 1$. Note that this discontinuity of the first derivative of the GS energy vanishes at the bi-critical point $(h,\gamma)=(1,0)$.
\\ To conclude, observe that the second order QPTs of the non-frustrated case \cite{Franchini_2017} survive when FBC are applied, see Eqs. \eqref{eq::E0} and \eqref{def DeltaE}. This is because they are bulk (\textit{i.e.} extensive) QPTs \cite{sachdev_2011} and, as a consequence, they are not sensitive to the choice of boundary conditions. At these critical lines the dispersion relation of Eq. \eqref{epsilon} vanishes for some values of $q$ and is relativistic (except the bi-critical point), differently from what happens at $h=0$, where $\epsilon(q)$ is Galilean around its minimum
\begin{equation}
    \epsilon(q)|_{h=0} \approx 1 + \frac{\gamma^2-1}{2}\ q^2\,,
\end{equation}
meaning that the energy gap between the ground state and the lightest excitation goes like $\delta \epsilon (q) \simeq \frac{\gamma^2-1}{2} q^2$.

\section{Second order b-QPTs}
\label{sec:secondorder}
In Sec. \ref{sec::Signatures of topological frustration} it has been shown that the critical line $\gamma^*(N,h)$ divides the region $0\leq h<1$ of the first quarter of the parameter space into two subregions with completely different properties. Crossing this line the GS changes from a gapless non degenerate one, with negative $z$-parity and vanishing momentum, to a doubly degenerate one with two states with the same $z$-parity but opposite, non vanishing, momenta. Because of the quantization of momenta -- see Eq. \eqref{Gamma sets} -- and the fact that the minima of $\epsilon(q)$ continuously change when moving $\gamma$ for $\gamma<\gamma^*(h)$, increasing $N$ the number of level crossings between two-fold degenerate manifolds with opposite $z$-parities and different non vanishing momenta increases. The gap between these alternating manifolds closes exponentially as $N$ increases, giving rise to a four times degenerate gapless region. Relevantly, in the thermodynamic limit the crossover between ground states with opposite parities becomes continuous, giving rise to an extreme case of orthogonality catastrophe \cite{Anderson1,Anderson2} which is a typical feature of one-dimensional models with continuous symmetries \cite{Catalano22}, such as the XXZ chain \cite{Franchini_2017}. Thus, a natural question is whether crossing  $\gamma^*(h)$ induces a b-QPT. In the following Subsection we will analytically prove that this is the case, and in Sec. \ref{sec::numerical} we will provide further numerical evidences that corroborate the correctness of our proof.
\subsection{Analytical approach}\label{sec::analytical}
By using the same notation introduced in Section \ref{sec:firstorder}, we have that
\begin{equation}\label{first derivative}
    \pdv{\Delta E}{\gamma} = \begin{cases}
    \frac{(1-\gamma^2)^2 -h^2}{(1-\gamma^2)^{3/2} \sqrt{1-h^2-\gamma^2}} &\mbox{if}\quad 0<\gamma <\gamma^*(h)
    \\[1.5 ex]
    0  &\mbox{if}\quad \gamma \geq\gamma^*(h)
    \end{cases}\,.
\end{equation}
\begin{figure}[h!]
    \centering
    \includegraphics[scale=1]{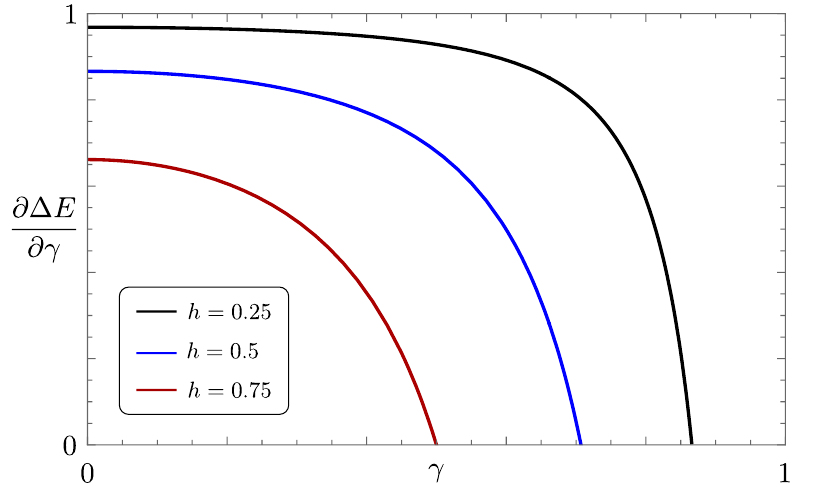}
    \caption{Plot of $\pdv{\Delta E}{\gamma}$ as a function of $\gamma$ for different values of $h$: $h=0.25$ (black),  $h=0.5$ (blue) and $h=0.75$ (red).}
    \label{fig dDeltaE J=1 intorno di gamma=0}
\end{figure}
Figure \ref{fig dDeltaE J=1 intorno di gamma=0} shows a plot of Eq. \eqref{first derivative} as a function of $\gamma$. The {\em first} derivative of $\frac{\partial \Delta E}{\partial\gamma}$ has a discontinuity at $\gamma=\gamma^{*}(h)$ and therefore the {\em second} derivative of $E$ must have a jump discontinuity for $\gamma = \gamma^*(h)$.  In particular, we observe that
\begin{equation}\label{b-QPT parabola}
    \lim_{\gamma\to (\gamma^*)^+}\pdv[2]{E}{\gamma}-\lim_{\gamma\to (\gamma^*)^-}\pdv[2]{E}{\gamma} = \frac{4}{h},
\end{equation}
from which we arrive at the key result that the system undergoes a {\em second order b-QPT} at $\gamma^*(h)$. Note that the divergence of the discontinuity for $h\rightarrow 0$ seems compatible with the emergence of the first order b-QPT discussed previously.\\
It should be noted that this discontinuity was not found in \cite{Catalano22}, where the same model was analyzed. While here we took the thermodynamic limit first and studied the energy discontinuity across the point $\gamma = \gamma^*$, in \cite{Catalano22} the authors calculated the finite-size energy jump and performed a finite-size scaling analysis. In the latter way, for any finite $N$ it is always possible to select a neighbor of the transition line where the momentum of ground state does not change with $\gamma$ and $N$ because of momentum quantization (see Eq. \eqref{Gamma sets}). Calculated in this way, the energy jump vanishes faster with large $N$, but it is not clear if the thermodynamic limit can be safely taken, since in this limit the constant momentum neighbor shrinks to zero. Thus we will support our claim that the procedure employed in this work is correct through numerical analysis in the next Section. Before doing so, it is important to stress here that in sharp contrast from all the other critical points studied in Section \ref{sec:firstorder}, 
the dispersion relation for the lightest excitation at the critical parabola $\gamma^{*}(h)$ has the following form
\begin{equation}
    \epsilon(q)|_{\gamma^*} \approx 1-h + \frac{h}{8(1-h)}\ q^4\,,
\end{equation}
indicating an energy gap of $\delta \epsilon (q)|_{\gamma^*} \simeq \frac{h}{8(1-h)} q^4$ which is thus neither Galilean nor relativistic, but quartic. To the best of our knowledge, this is the first case of second order QPT characterized by such a dispersion relation. In the scaling limit, quartic dispersion relations would be generated by high derivative or long-range field theories, but these terms are typically irrelevant and their contributions are usually unobservable corrections to the leading terms. In this case, this dispersion relation indicate a highly non-linear behavior of the single topological excitation injected in the GS by frustration \cite{torre2023longrange}, which is likely a quantum analog to the quartic solitons \cite{quartic1,quartic2,quartic3,quartic4}.

\subsection{Numerical analysis}
\label{sec::numerical}
In this Subsection we support our previous theoretical results with a numerical approach that confirms the existence of a second order b-QPT along the critical parabola $\gamma^*(h)$.\\ 

\noindent The second partial derivative of the GS energy $E(h,\gamma;N)$ with respect to $\gamma$ for finite odd $N$ has been evaluated employing a fourth--order "stencil" \cite{as}
\begin{align}\label{def_num_der}
    \pdv[2]{E(h,\gamma;N)}{\gamma} \simeq \frac{-E(\gamma+2\delta\gamma)+16E(\gamma+\delta\gamma)-30E(\gamma)+16E(\gamma-\delta\gamma)-E(\gamma-2\delta\gamma)}{12\delta\gamma^2},
\end{align}
where, for brevity, we have denoted $E(\gamma)=E(h,\gamma; N)$ to the right hand side of the above equation. The truncation error of Eq. (\ref{def_num_der}) is $\order{(\delta\gamma)^4}$. This estimate of $\pdv[2]{E(h,\gamma;N)}{\gamma}$ has shown to be numerically stable as long as $\delta\gamma<\frac{2\pi}{N}$, see App. \ref{app::numstab}.\\

\begin{figure}[h!]
    \centering
    \includegraphics[scale=0.55]{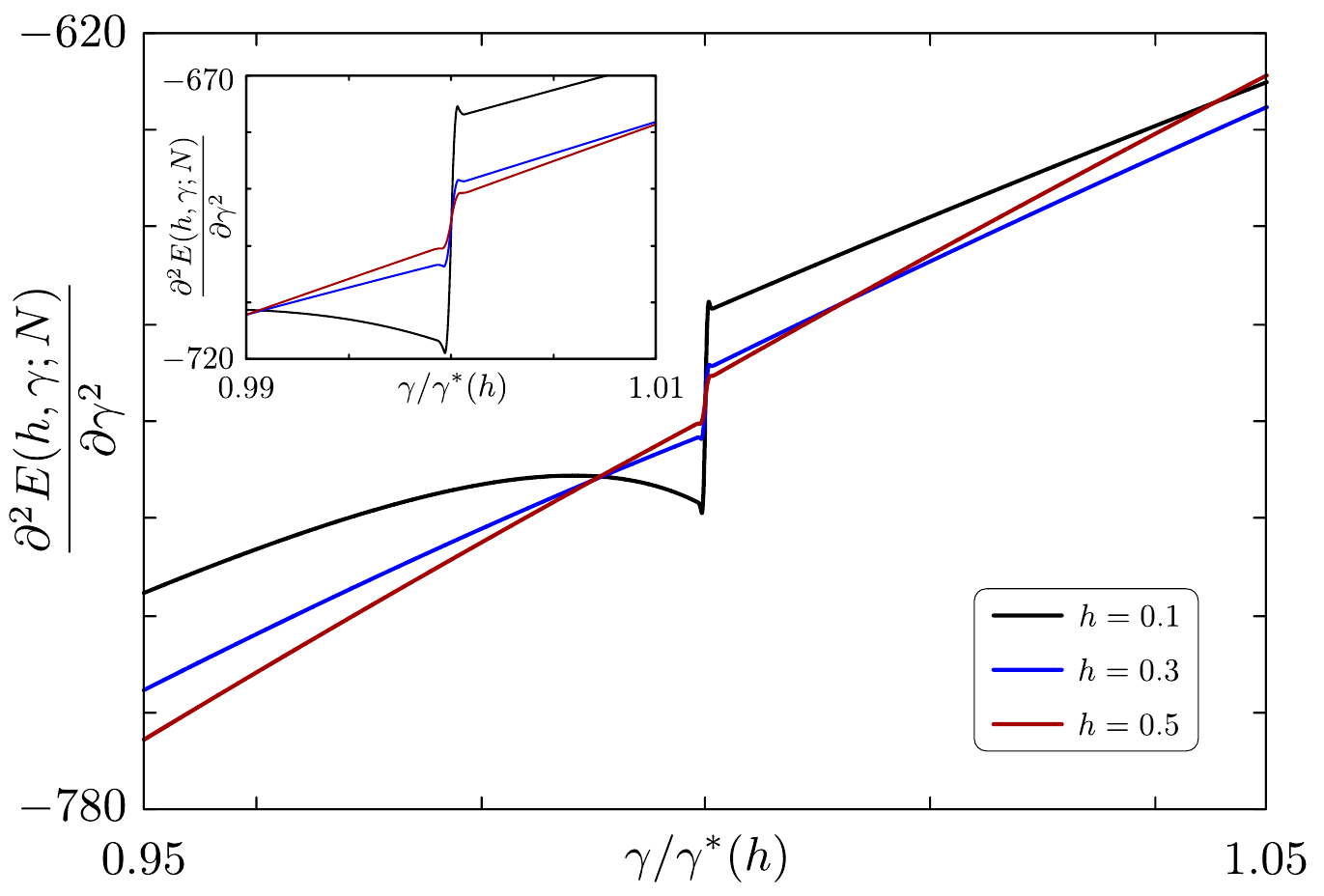}
    \caption{Plot of $\frac{\partial^2 E(h,\gamma;N)}{\partial\gamma^2}$ as a function of $\gamma/\gamma^{*}(h)$  for different values of $h$: $h=0.1$ (black), $h=0.3$ (blue), $h=0.5$ (red). The top left inset shows a zoom centered about the abrupt jump. Here $N=10001$ and note that the curves have been offset in order to vertically align the center of the jump region.}
    \label{fig::overview}
\end{figure}

\noindent Figure \ref{fig::overview} shows the behaviour of $\pdv[2]{E(h,\gamma;N)}{\gamma}$ as a function of $\gamma/\gamma^{*}(h)$ at finite odd $N$ for three different values of $h$. As is clear, a sudden jump is found around $\gamma=\gamma^{*}(h)$. Crucially, this jump increases as $h$ decreases. The inset of Fig. \ref{fig::overview} shows a zoom in a narrow window of $\gamma$ around $\gamma^{*}(h)$: clearly, the curves exhibit a local minimum (for $\gamma=\bar{\gamma}_{-}\lesssim\gamma^{*}(h)$) followed by a sudden jump and and a local maximum (for $\gamma=\bar{\gamma}_{+}\gtrsim\gamma^{*}(h)$), a behaviour closely reminiscent of the Gibbs phenomenon \cite{Hewitt1979TheGP, Gottlieb}. As $N$ is increased, the distance $\Delta\gamma=\bar{\gamma}_+-\bar{\gamma}_-$ is found to scale as $1/N$, see App. \ref{app::numstab}. This, together with the fact that the height of the jump is stable increasing $N$ -- see below and App. \ref{app::numstab} -- strongly suggests that in the thermodynamic limit $\frac{\partial^2 E(h,\gamma;N)}{\partial\gamma^2}$ indeed develops a discontinuity.\\ 
\begin{figure}[h!]
    \centering
    \includegraphics[scale=0.55]{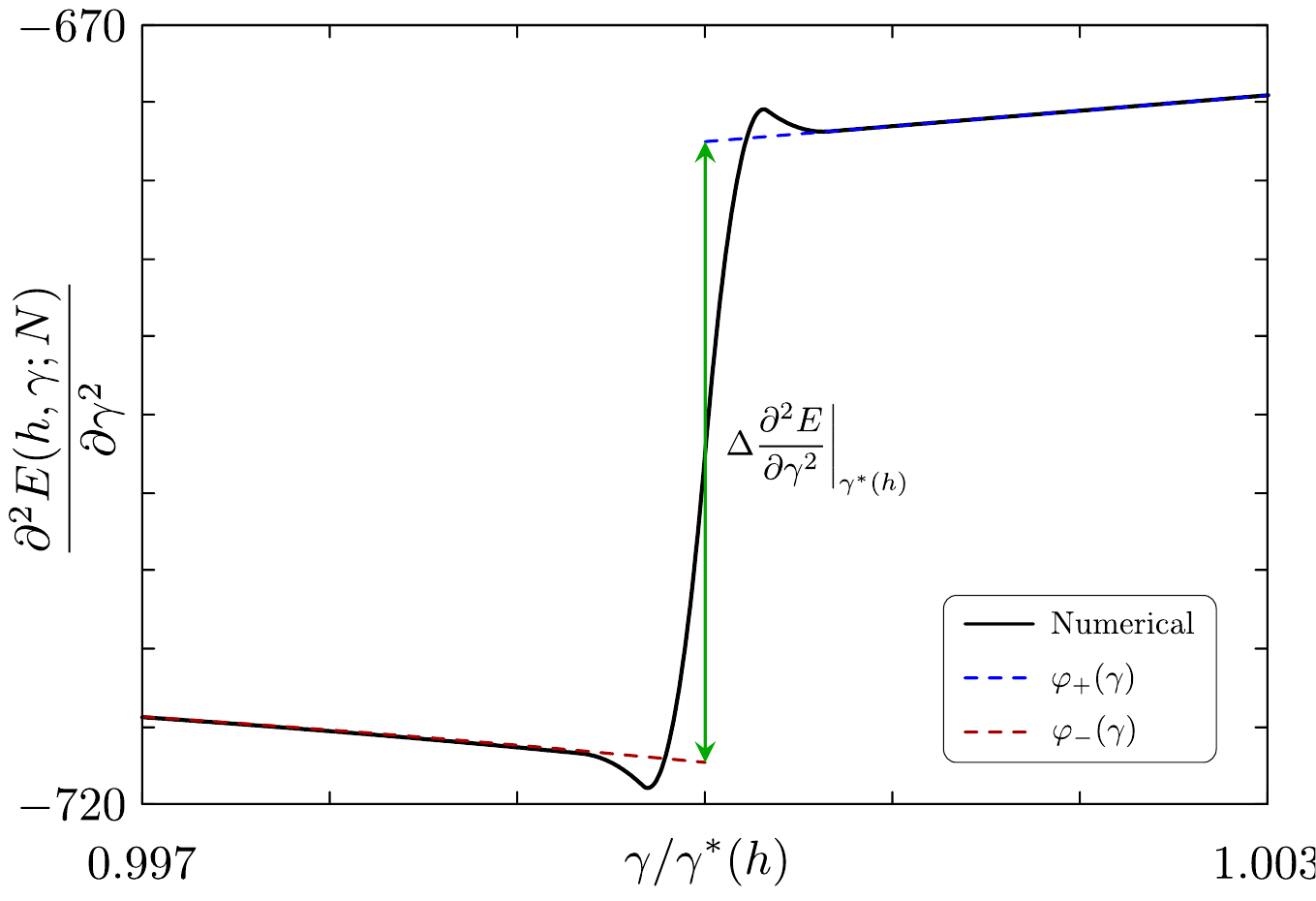}
    \caption{Plots of $\frac{\partial^2 E(h,\gamma;N)}{\partial\gamma^2}$ as a function of $\gamma/\gamma^{*}(h)$ (black solid line) and of the extrapolating functions $\varphi_{-}(\gamma)$ (red dashed line) and $\varphi_{+}(\gamma)$ (blue dashed line). Here, $h=0.1$ and $N=10001$.}
    \label{fig::extrapolating_procedure}
\end{figure}
To obtain the value of this discontinuity without ambiguity, we extrapolate the jump of $\pdv[2]{E(h,\gamma;N)}{\gamma}$ around $\gamma^{*}(h)$ at finite $N$ introducing two third--order polynomial approximations $\varphi_{-}(\gamma)$ and $\varphi_{+}(\gamma)$, obtained fitting the numerical data for $\gamma<\gamma_{-}$ and $\gamma>\gamma_{+}$. We then define the jump of the second derivative at finite $N$ as
\begin{equation}\label{def extrapolated gap}
    \Delta \pdv[2]{E}{\gamma}\bigg|_{\gamma^*(h)} \equiv \varphi_+(\gamma^*(h))-\varphi_-(\gamma^*(h)).
\end{equation}
For large $N$, this quantity shows to be independent of $N$ -- see App. \ref{app::numstab} -- providing evidence to the fact that the jump is stable in the thermodynamic limit. The procedure is illustrated in Fig. \ref{fig::extrapolating_procedure}, which shows the second derivative of the GS energy and the approximating functions $\varphi_{-}(\gamma)$ and $\varphi_{+}(\gamma)$ (red and blue dashed curves respectively) as a function of $\gamma/\gamma^{*}(h)$. .\\
\begin{figure}[ht!]
    \centering
    \includegraphics[scale=0.55]{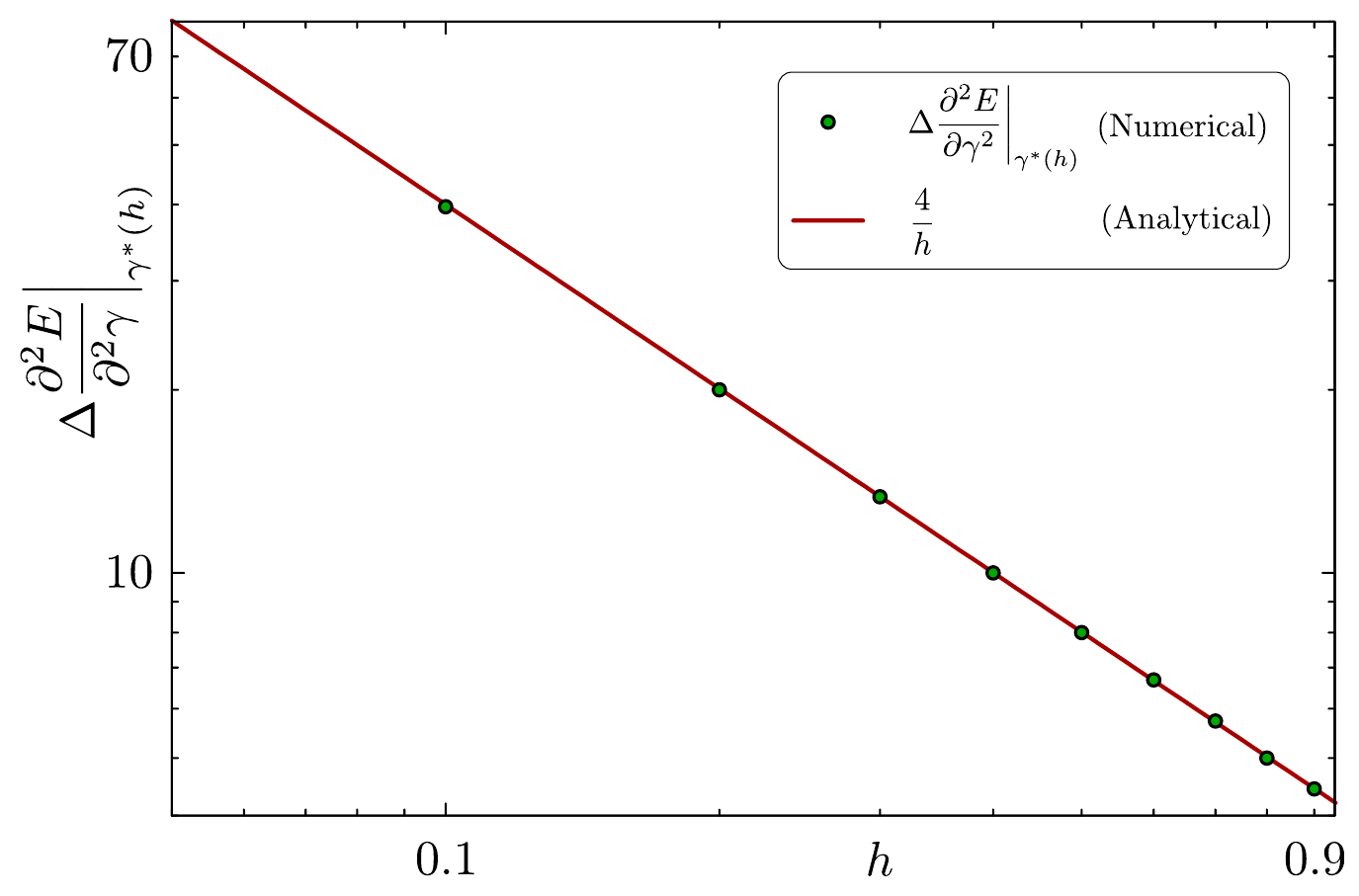}
    \caption{Plots of the numerically evaluated gap $\Delta \pdv[2]{E}{\gamma}\big|_{\gamma^*(h)}$ (defined by Eq. \eqref{def extrapolated gap}) as a function of $h$ (green dots) and of the analytical result $4/h$. Here, $N=100001$.}
    \label{fig::scaling of the gap}
\end{figure}
\noindent At last, we show in Fig. \ref{fig::scaling of the gap} the value of the jump of the second derivative of the GS energy obtained numerically as green dots, compared with the analytical result $4/h$ obtained in Eq. \eqref{b-QPT parabola}. The agreement is excellent, which confirms the robustness of the analytical calculations performed in Subsec. \ref{sec::analytical}.
\section{Conclusions}\label{sec::conclusion}
We have shown that TF can deeply modify the zero temperature phase diagram of a one-dimensional quantum spin-$1/2$ system with discrete symmetries by inducing new b-QPTs, both of the first and of the second order. In particular, we focused on the example of the XY chain in an external magnetic field, possibly the simplest non-trivial integrable model. Without frustration, this model is characterized by two QPTs: one at $\gamma=0$, $h\leq 1$ (belonging to the universality class with conformal charge $c=1$) which separates two ordered phases and the other one at $h=1$ ($c=1/2$ CFT) that divides the ordered phase from the disordered (\textit{i.e.} paramagnetic) one. These two lines meet at the bi-critical point $(h,\gamma)=(1,0)$, which is non-conformal. When FBC are imposed, first order b-QPTs are induced at the conformal lines and at $h=0$, $\gamma\geq 1$, where the dispersion relation is Galilean. Furthermore, TF generates a second order b-QPT in correspondence of the parabola $\gamma = \sqrt{1-h}$ which separates a non-degenerate gapless region with definite $z$-parity and zero momentum from a four times degenerate gapless region with non-vanishing momenta and characterized by an extreme case of orthogonality catastrophe (which is a typical feature of models with continuous symmetries).\\ 
\begin{figure}[h]
    \centering
    \includegraphics[width=0.785\textwidth]{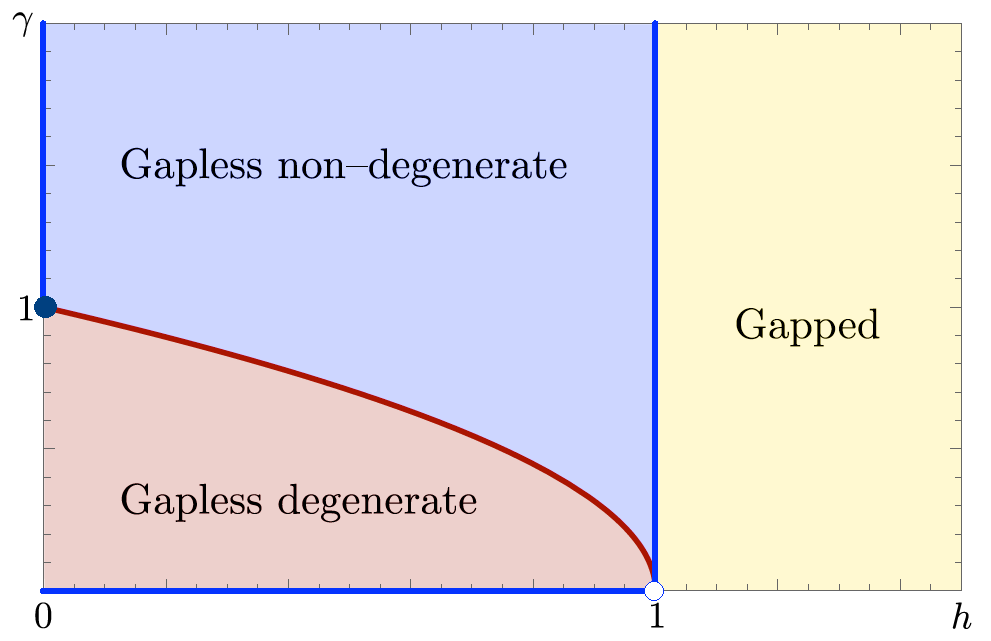}
    \caption{Zero temperature phase diagram of the frustrated XY chain. The blue (red) bold lines represent the first (second) order boundary QPT induced by the imposition of FBC. The empty circle at (1,0) stands for the absence of boundary quantum phase transitions and the filled circle at (0,1) indicates the presence of a first order boundary quantum phase transition.}
    \label{fig::phase diagram}
\end{figure}
\noindent Figure \ref{fig::phase diagram} resumes all the QPTs induced by FBC in the XY spin-$1/2$ chain.\\ 

\noindent Interestingly, to the best of our knowledge we also came across the first example of second order  b-QPT characterized by a dispersion relation which is neither relativistic nor Galilean but quartic. The correctness of our analytical calculations is supported by a thorough numerical analysis.\\ 

\noindent In conclusion, our work lays the foundation for the construction of the zero temperature phase diagram of the topologically frustrated XY chain in a transverse magnetic field. A natural question that emerges is what are the new order parameters that characterize the phases separated by the new QPTs found in this work. This point is still unanswered but we guess that the new phases could be characterized by topological order parameters. This open problem will surely be the subject of future investigations. 

\section*{Acknowledgements}
D.S.S. acknowledges Davide Rovere and Dario Ferraro for discussions.

\paragraph{Funding information}
N.T.Z. acknowledges the funding through the NextGenerationEu Curiosity Driven Project ``Understanding even-odd criticality''.  N.T.Z. and M.S. acknowledge the funding through the ``Non-reciprocal supercurrent and topological transitions in hybrid Nb-InSb nanoflags'' project (Prot. 2022PH852L) in the framework of PRIN 2022 initiative of the Italian Ministry of University (MUR) for the National Research Program (PNR). F.F. acknowledges support from the Croatian Science Foundation (HrZZ) Project No. IP–2019–4–3321. A.G.C. is supported by the MOQS ITN programme, a European Union’s Horizon 2020 research and innovation program under the MSCA grant agreement number 955479. F.C. would like to acknowledge the contribution of the European Union-NextGenerationEU through the ”Solid State Quantum Batteries: Characterization and Optimization” (SoS-QuBa) project, in the framework of the PRIN 2022 initiative of the Italian Ministry of University (MUR) for the National Research Program (PNR).

\begin{appendix}

\section{Exact diagonalization}
\label{app:diagonalization}
We introduce the Wigner-Jordan transformation \cite{Jordan:1928wi}, in which the spin operators are mapped onto non-local spinless fermions. One has
\begin{align}\label{def trasf Wigner-Jordan}
    \sigma_j^+ \equiv \exp{i\pi \sum_{l=1}^{j-1}\psi_l^\dagger \psi_l}\psi_j \qquad j=1,...,N,
\end{align}
with $\sigma_j^\pm=\sigma_j^x\pm i\sigma_j^y$. The operators $\psi_l$ satisfy the anti--commutation relations
\begin{equation}\label{ferm}
\{\psi_l^\dagger,\psi_{l'}\}=\delta_{l,l'} \quad,\qquad \{\psi_l,\psi_{l'}\}=0
\end{equation}
with $l$, $l'$ running over the sites of the chain and $\delta_{l,l'}$ the Kronecker symbol.\\
With respect to the new basis of operators, the Hamiltonian $H$ reads 
\begin{align}
    H = &\frac{J}{2}\sum_{j=1}^{N-1}\left(\psi_{j+1}^\dagger \psi_j + \psi_j^\dagger \psi_{j+1} + \gamma \psi_{j+1}\psi_j + \gamma \psi_j^\dagger \psi_{j+1}^\dagger\right) +\frac{JhN}{2} -Jh\sum_{j=1}^N \psi_j^\dagger \psi_j + \nonumber\\
    &-\frac{J}{2}\Pi^z\left(\psi_1^\dagger \psi_N + \psi_N^\dagger \psi_1  +\gamma \psi_1\psi_N + \gamma \psi_N^\dagger \psi_1^\dagger\right),\label{H xy in fne di psi}
\end{align}
with
\begin{equation}
    \Pi^\alpha \equiv \bigotimes_{l=1}^N \sigma_l^\alpha ,
\end{equation}
the parity operator. We note that the fermionic form of the Hamiltonian is highly non-local and non-\textcolor{black}{quadratic}. However, since
\begin{equation}\label{z-parity simmetry}
    [H,\Pi^z] =0.
\end{equation}
one can decompose the Hamiltonian as
\begin{equation}\label{z parity constraint}
    H = \frac{1+\Pi^z}{2} H^+ \frac{1+\Pi^z}{2} + \frac{1-\Pi^z}{2} H^- \frac{1-\Pi^z}{2},
\end{equation}
where \textcolor{black}{$H^+$ and $H^-$ are quadratic (and hence writable in a free-fermionic form)}:
\begin{align}
    H^\pm = &\frac{J}{2}\sum_{j=1}^{N-1}(\psi_{j+1}^\dagger \psi_j + \psi_j^\dagger \psi_{j+1} + \gamma \psi_{j+1}\psi_j + \gamma \psi_j^\dagger \psi_{j+1}^\dagger) +\frac{JhN}{2} -Jh\sum_{j=1}^N \psi_j^\dagger \psi_j + \nonumber\\
    &\mp\frac{J}{2}(\psi_1^\dagger \psi_N + \psi_N^\dagger \psi_1  +\gamma \psi_1\psi_N + \gamma \psi_N^\dagger \psi_1^\dagger)\\
    \equiv& \frac{J}{2}\sum_{j=1}^N(\psi_{j+1}^{(\pm)\dagger}\psi^{(\pm)}_j + \psi_j^{(\pm)\dagger}\psi^{(\pm)}_{j+1} +\gamma\psi^{(\pm)}_{j+1}\psi^{(\pm)}_j + \gamma\psi_j^{(\pm)\dagger}\psi_{j+1}^{(\pm)\dagger}   -2h\psi_j^{(\pm)\dagger}\psi^{(\pm)}_j)+\frac{JhN}{2}\label{H^pm(psi^pm)}\,,
\end{align}
\textcolor{black}{with}
\begin{align}\label{def psi^pm}
    \begin{cases}
    \psi_j^{(\pm)} \equiv \psi_j  \\[1.6 ex]
    \psi^{(\pm)}_{j+N} \equiv \mp\psi^{(\pm)}_{j}
    \end{cases}
    \qquad j= 1,...,N.
\end{align}
We are now in the position of diagonalizing the Hamiltonian. We switch to Fourier space, introducing operators $\tilde{\psi}_{q}$ via the relation
\begin{equation}\label{def trasf fourier fermioni di Wigner-Jordan}
    \psi_j^{(\pm)} \equiv \frac{1}{\sqrt{N}} e^{i\frac{\pi}{4}}\sum_{q\in \Gamma^\pm}e^{iqj}\tilde{\psi}_q,
\end{equation}
where
\begin{equation}\label{def Gamma^pm}
    \Gamma^+ \equiv \left\{\frac{2\pi}{N}\left(k+\frac{1}{2}\right)\right\} \qquad \Gamma^- \equiv \left\{\frac{2\pi}{N}k\right\}\qquad k=0,...,N-1.
\end{equation}
In Fourier space one finds
\begin{align}
    H^{\pm} &= -J\sum_{q\in \Gamma^\pm}(h-\cos q)\tilde{\psi}_q^\dagger \tilde{\psi}_q -\frac{J\gamma}{2}\sum_{q\in\Gamma^\pm}\sin q (\tilde{\psi}_q \tilde{\psi}_{2\pi-q} + \tilde{\psi}^{\dagger}_{2\pi-q} \tilde{\psi}_q^\dagger) + \frac{JhN}{2}\nonumber\\
    &= \frac{J}{2}\sum_{q\in\Gamma^\pm}\begin{pmatrix}
    \tilde{\psi}_q^{\dagger} & \tilde{\psi}_{2\pi-q}
    \end{pmatrix}
    \begin{pmatrix}
    -h+\cos q & \gamma \sin q\\
    \gamma \sin q & h-\cos q
    \end{pmatrix}
    \begin{pmatrix}
    \tilde{\psi}_q\\ \tilde{\psi}^{\dagger}_{2\pi-q}
    \end{pmatrix}.
\end{align}
We now define new operators $\chi_{q}$ via the Bogoljubov rotation
\begin{equation}\label{rotazione Bogoliubov}
    \begin{pmatrix}
    \tilde{\psi}_{q}\\[1ex] \tilde{\psi}^{\dagger}_{2\pi-q}
    \end{pmatrix} \equiv
    \begin{pmatrix}
    \cos\theta_q & \sin \theta_q\\[1ex]
    -\sin \theta_q & \cos\theta_q
    \end{pmatrix}
    \begin{pmatrix}
    \chi_{q}\\[1ex] \chi^{\dagger}_{2\pi-q}
    \end{pmatrix},
\end{equation}
where the angle $\theta_q$ obeys the constraints
\begin{equation}
    \theta_{0,\pi} \equiv 0,
\end{equation}
\begin{equation}
    \begin{cases}
        \sin \theta_{2\pi-q} = -\sin\theta_q\\
        \cos\theta_{2\pi-q} = \cos\theta_q.
    \end{cases}
\end{equation}
In terms of these new operators the Hamiltonian reads
\begin{align}
    H^{\pm}  =& \frac{J}{2}\sum_{q\in\Gamma^\pm}\begin{pmatrix}
    \chi_q^{\dagger} & \chi_{2\pi-q}
    \end{pmatrix}
    \Tilde{H}
    \begin{pmatrix}
    \chi_q\\ \chi^{\dagger}_{2\pi-q}
    \end{pmatrix}
\end{align}
with
\begin{equation}
    \Tilde{H} = \begin{pmatrix}
    (-h+\cos q)\cos(2\theta_q)-\gamma\sin q \sin (2\theta_q) & \gamma \cos(2\theta_q)\sin q + (-h+\cos q)\sin(2\theta_q)\\
    \gamma \cos(2\theta_q)\sin q + (-h+\cos q)\sin(2\theta_q) & (h-\cos q)\cos(2\theta_q)+\gamma\sin q \sin (2\theta_q)
    \end{pmatrix}.
\end{equation}
This form brings us to the final step of the diagonalization. Here, one needs to distinguish between the most common the ferromagnetic case with $J<0$ and the antiferromagnetic case $J>0$, the one that shows frustration (which is the focus of our work).
\subsection{The ferromagnetic case ($J<0$)}
In this extensively studied case \cite{Franchini_2017,porta} it is convenient to chose $\theta_q$ such that
\begin{equation}\label{def angolo di Bogoliubov}
    e^{i2\theta_q} = \frac{h-\cos q  +i\gamma \sin q}{\sqrt{(h-\cos q)^2 + \gamma^2 \sin^2q}} \qquad q\neq 0,\pi.
\end{equation}
Furthermore, we introduce the dispersion relation
\begin{equation}
    \epsilon(q)\equiv \sqrt{(h-\cos q)^2 + \gamma^2 \sin^2q}.
\end{equation}
Note that
\begin{align}
    &\epsilon(\pi) 
    = h+1 \qquad\mbox{if}\quad h\geq 0\\
    &\epsilon(0) 
    = \begin{cases}
    h-1 &\quad\mbox{if}\quad h>1\\
    -h+1&\quad\mbox{if}\quad 0\leq h<1\,.
    \end{cases}
\end{align}
From Eq. \eqref{def angolo di Bogoliubov} it follows that
\begin{align}
    H^+ =& -J\sum_{q\in\Gamma^+}\epsilon(q)\left(\chi_q^\dagger\chi_q-\frac{1}{2}\right)\label{H^+ferro}\\
    H^- =& \begin{cases}
    -J\sum_{q\in \Gamma^-}\epsilon(q)\left(\chi_q^{\dagger} \chi_q-\frac{1}{2}\right) &\quad\mbox{if}\quad  h>1\\[1 ex]
    -J\sum_{q\in \Gamma^-\setminus\{0\}}\epsilon(q)\left(\chi_q^{\dagger} \chi_q-\frac{1}{2}\right) +J\epsilon(0)\left(\chi_0^{\dagger} \chi_0-\frac{1}{2}\right) &\quad\mbox{if}\quad 0\leq h<1 \label{H^-ferro}
    \end{cases}
\end{align}
independently from the parity of $N$.
\subsection{The antiferromagnetic case ($J>0)$}
Here $\theta_q$ is chosen such that
\begin{equation}\label{def angolo di Bogoliubov antiferro}
    e^{i2\theta_q} = \frac{-h+\cos q  -i\gamma \sin q}{\sqrt{(h-\cos q)^2 + \gamma^2 \sin^2q}} \qquad q\neq 0,\pi.
\end{equation}
Now one needs \textcolor{black}{to distinguish} between the even $N$ case and the odd one. \\For even $N$ one has
\begin{align}
    H^+ =& 
    J\sum_{q\in\Gamma^+}\epsilon(q)\left(\chi_q^\dagger \chi_q-\frac{1}{2}\right)\label{H^+ even}\\
    H^- =& \begin{cases}
    J\sum_{q\in \Gamma^-\setminus\{0,\pi\}}\epsilon(q)\left(\chi_q^{\dagger} \chi_q-\frac{1}{2}\right)-J\epsilon(\pi)\left(\chi_\pi^{\dagger} \chi_\pi-\frac{1}{2}\right)-J\epsilon(0)\left(\chi_0^{\dagger} \chi_0-\frac{1}{2}\right) &\quad\mbox{if}\quad  h>1\\[1 ex]
    J\sum_{q\in \Gamma^-\setminus\{\pi\}}\epsilon(q)\left(\chi_q^{\dagger} \chi_q-\frac{1}{2}\right) -J\epsilon(\pi)\left(\chi_\pi^{\dagger} \chi_\pi-\frac{1}{2}\right) &\quad\mbox{if}\quad 0\leq h<1 .
    \end{cases}\label{H^- even}
\end{align}
For odd $N$, which is the frustrated case and represents the main focus of this work, one has
\begin{align}
    H^+ =& J\sum_{q\in\Gamma^+\setminus\{\pi\}}\epsilon(q)\left(\chi_q^\dagger \chi_q-\frac{1}{2}\right)-J\epsilon(\pi)\left(\chi_\pi^{\dagger} \chi_\pi-\frac{1}{2}\right)\label{H^+ odd}\\
    H^- =& \begin{cases}
    J\sum_{q\in \Gamma^-\setminus\{0\}}\epsilon(q)\left(\chi_q^{\dagger} \chi_q-\frac{1}{2}\right)-J\epsilon(0)\left(\chi_0^{\dagger} \chi_0-\frac{1}{2}\right) &\quad\mbox{if}\quad  h>1\\[1 ex]
    J\sum_{q\in \Gamma^-}\epsilon(q)\left(\chi_q^{\dagger} \chi_q-\frac{1}{2}\right)  &\quad\mbox{if}\quad 0\leq h<1 
    \end{cases}\label{H^- odd}.
\end{align}
\section{Stability and scaling of the numerical results}
\label{app::numstab}
In this Section we briefly comment on the stability of the numerical method defined in Eq. \eqref{def_num_der} and on its scaling properties against the number of particles $N$.\\

\noindent Concerning the stability of the fourth order stencil of Eq. \eqref{def_num_der}, a sufficiently small value for $\delta\gamma$ has to be employed. A sensible order of magnitude for $\delta\gamma$ is given by the smallest typical energy scale of the system, i.e. $\delta k = 2\pi/N$.
\begin{figure}[h!]
    \centering
    \includegraphics[scale=0.5]{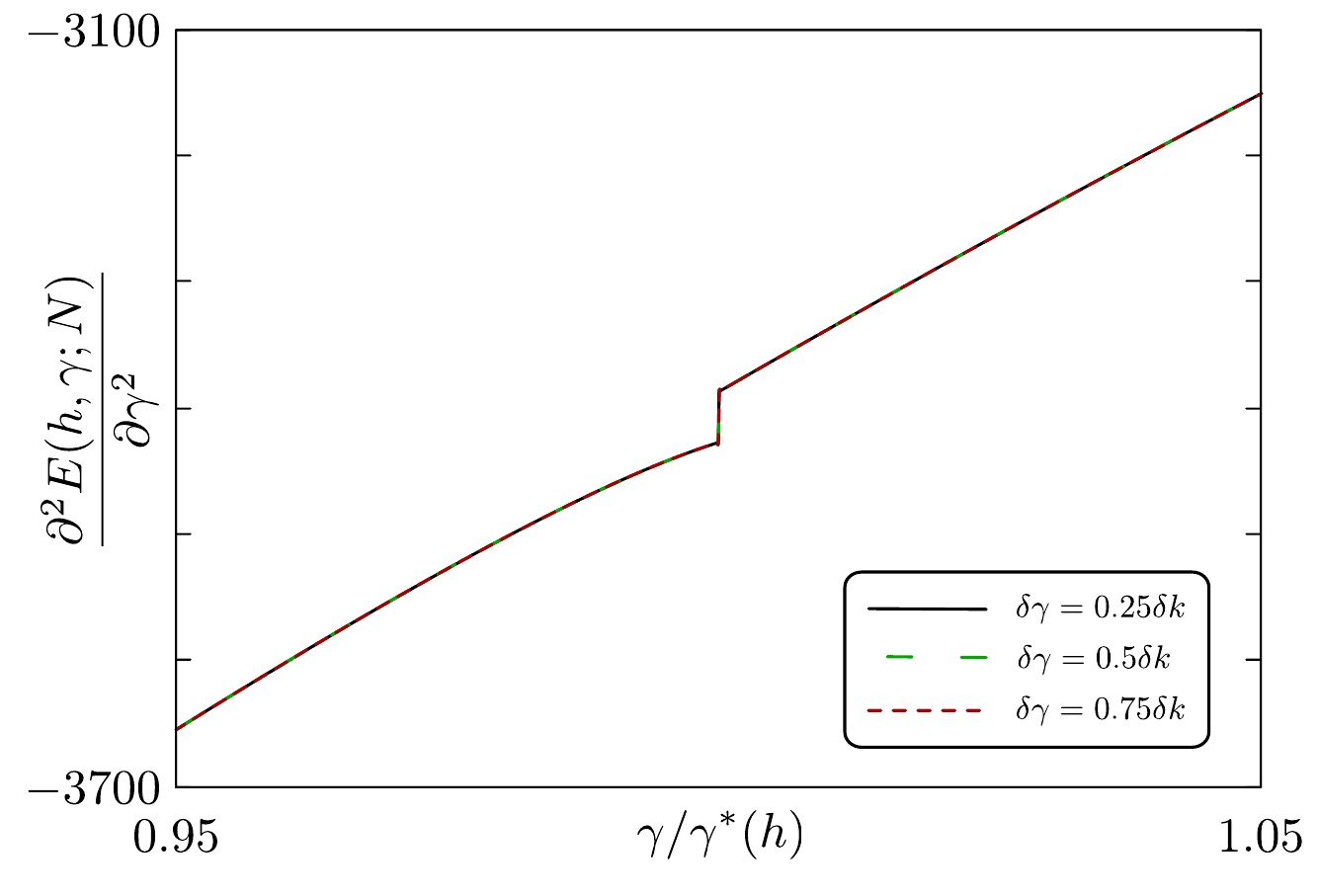}
    \caption{Plot of $\frac{\partial^2 E(h,\gamma;N)}{\partial\gamma^2}$ as a function of $\gamma/\gamma^{*}(h)$ for three different values of $\delta\gamma$: $\delta\gamma=0.25\delta k$ (black), $\delta\gamma=0.5\delta k$ (green dashed), $\delta\gamma=0.75\delta k$ (red dotted). Here, $h=0.1$ and $N=50001$.}
    \label{fig::derivative-stability}
\end{figure}
Figure \ref{fig::derivative-stability} shows a plot of $\frac{\partial^2 E(h,\gamma;N)}{\partial\gamma^2}$ as a function of $\gamma/\gamma^{*}$ for three different choices of $\delta\gamma<\delta k$. As can be seen, all curves collapse ontop of each other, signalling that the derivative is stable with respect to the choice of $\delta\gamma<\delta k$. All the \textcolor{black}{subsequent} numerical analyses have been performed employing $\delta\gamma=0.25\delta k$.\\
\begin{figure}[h!]
    \centering
    \includegraphics[scale=0.5]{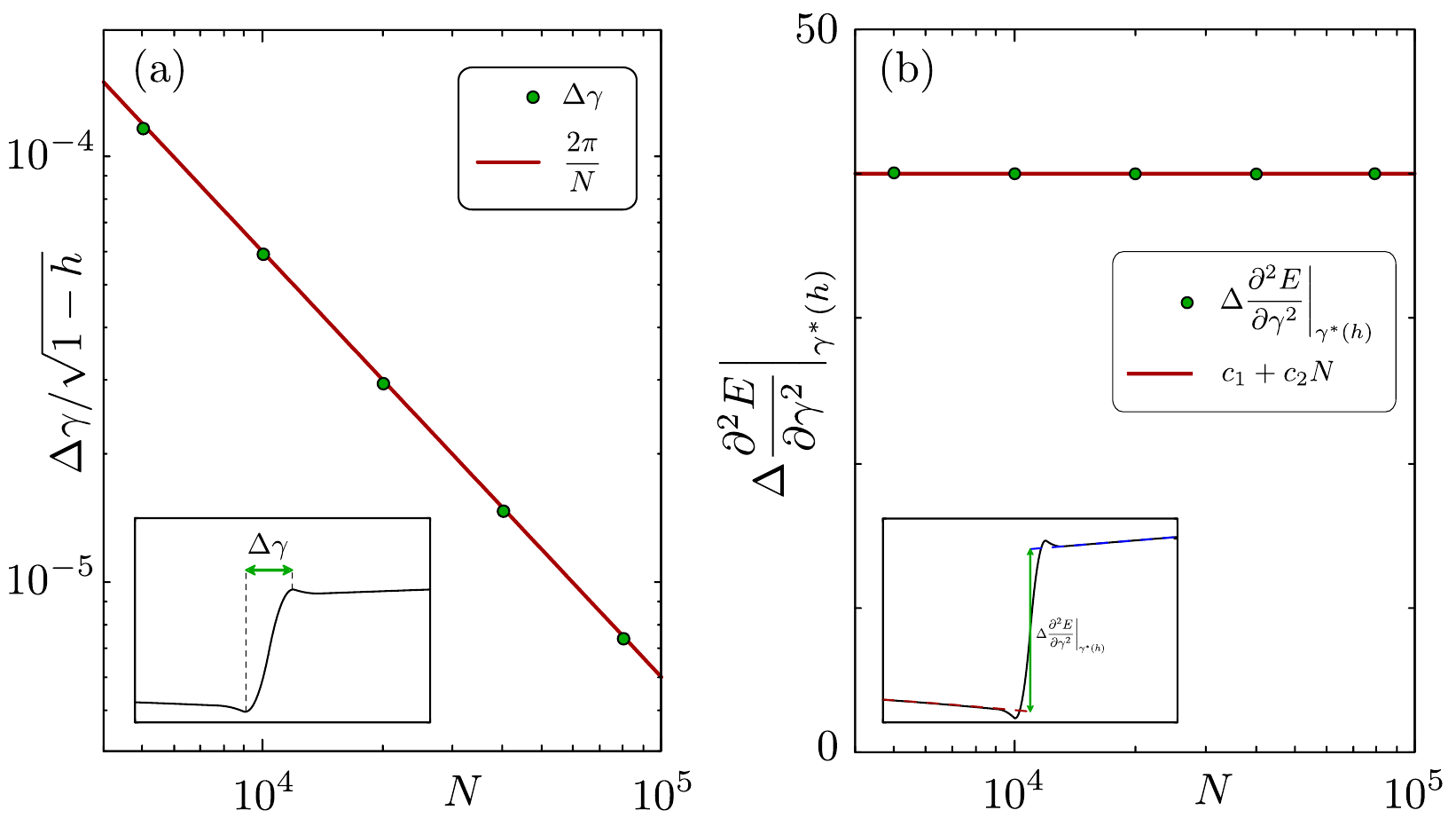}
    \caption{Panel (a): Plot of $\Delta\gamma/\sqrt{1-h}$ (green dots) and of the function $2\pi/N$ (red line) as a function of $N$. Panel (b): plot of $\left.\Delta\frac{\partial^2 E)}{\partial\gamma^2}\right|_{\gamma^{*}(h)}$ (green dots) and of the fitting function $c_1+c_2 N$ as a function of $N$. Here, $h=0.1$ and the fitting parameters have been found to be $c_1\approx39.97$ and $c_2\approx6.79\cdot10^{-11}$.}
    \label{fig::derivative-scaling}
\end{figure}

\noindent We now comment on the scaling properties of the numerical results with respect to $N$. As mentioned in Sec. \ref{sec::numerical}, the jump of the second derivative exhibits a under/overshooting behaviour around the jump, similar to the Gibbs phenomenon, characterized by a local minimum for $\gamma=\bar{\gamma}_{-}<\gamma^{*}(h)$ and a local maximum for $\gamma=\bar{\gamma}_{+}>\gamma^{*}(h)$. Via careful numerical analysis, we have found that the distance $\Delta\textcolor{black}{\gamma}=\bar{\gamma}_{+}-\bar{\gamma}_{-}$ scales with $N$ as
\begin{equation}
\Delta\gamma=\frac{2\pi}{N}\sqrt{1-h}.\label{eq::scal}
\end{equation}
An example of this behaviour is shown in Fig. \ref{fig::derivative-scaling}(a) for $h=0.1$, where the green dots represent numerical data obtained for different large values of odd $N$ and the red curve represents the scaling law of Eq. \eqref{eq::scal}. The agreement is excellent and the same level of accuracy of the above scaling law has been observed for all values of $h$. This suggests that in the thermodynamic limit $\frac{\partial^2 E(h,\gamma;N)}{\partial\gamma^2}$ indeed develops a discontinuity. The latter can be extrapolated, from the behaviour at finite $N$, via the procedure described in Sec. \ref{sec::numerical} that defines $\Delta\left.\frac{\partial^2 E}{\partial\gamma^2}\right|_{\gamma^{*}(h)}$ given in Eq. \eqref{def extrapolated gap}. Such quantity shows an excellent stability with respect to $N$. As an example, Fig. \ref{fig::derivative-stability}(b) shows it plotted as a function of large $N$ (green dots): the observed numerical independence of $N$ strongly supports the presence of a finite jump in the thermodynamic limit. To quantitatively assess the stability of this result we have fitted the numerical data of Fig. \ref{fig::derivative-scaling}(b) to a linear function of the form $c_1+c_2 N$. For the case of $h=0.1$ shown in the Figure, the fitting parameters are found to be $c_1\approx 39.97$ -- in excellent agreement with the theoretically found law $4/h$ --  and $c_2<10^{-11}$, which is a solid evidence of the insensitivity of the numerical results with respect to large $N$. Similar levels of robustness have been found for all values of $h$. This allows to conclude that our numerical results are thermodynamically robust.
\end{appendix}

\bibliography{SciPost_Example_BiBTeX_File.bib}

\begin{thebibliography}{10}
\providecommand{\url}[1]{\texttt{#1}}
\providecommand{\urlprefix}{URL }
\expandafter\ifx\csname urlstyle\endcsname\relax
  \providecommand{\doi}[1]{doi:\discretionary{}{}{}#1}\else
  \providecommand{\doi}{doi:\discretionary{}{}{}\begingroup \urlstyle{rm}\Url}\fi
\providecommand{\eprint}[2][]{\url{#2}}

\bibitem{Landau_Phase_Transitions}
L.~D. Landau,
\newblock \emph{{On the theory of phase transitions}},
\newblock Zh. Eksp. Teor. Fiz. \textbf{7}, 19 (1937),
\newblock \doi{10.1016/B978-0-08-010586-4.50034-1}.

\bibitem{LANDAU1980446}
L.~LANDAU and E.~LIFSHITZ,
\newblock \emph{Chapter xiv - phase transitions of the second kind and critical phenomena},
\newblock In L.~LANDAU and E.~LIFSHITZ, eds., \emph{Statistical Physics (Third Edition)}, pp. 446--516. Butterworth-Heinemann, Oxford, third edition edn.,
\newblock ISBN 978-0-08-057046-4,
\newblock \doi{https://doi.org/10.1016/B978-0-08-057046-4.50021-X} (1980).

\bibitem{GoldstoneWeinbergSalam}
J.~Goldstone, A.~Salam and S.~Weinberg,
\newblock \emph{Broken symmetries},
\newblock Phys. Rev. \textbf{127}, 965 (1962),
\newblock \doi{10.1103/PhysRev.127.965}.

\bibitem{Anderson_SSB}
P.~W. Anderson,
\newblock \emph{Plasmons, gauge invariance, and mass},
\newblock Phys. Rev. \textbf{130}, 439 (1963),
\newblock \doi{10.1103/PhysRev.130.439}.

\bibitem{Notes_SSB}
A.~J. Beekman, L.~Rademaker and J.~van Wezel,
\newblock \emph{{An introduction to spontaneous symmetry breaking}},
\newblock SciPost Phys. Lect. Notes p.~11 (2019),
\newblock \doi{10.21468/SciPostPhysLectNotes.11}.

\bibitem{Goldenfeld}
N.~Goldenfeld,
\newblock \emph{Lectures On Phase Transitions And The Renormalization Group},
\newblock Frontiers in physics. Westview Press,
\newblock ISBN 9780201554090,
\newblock \doi{10.1201/9780429493492} (1992).

\bibitem{sachdev_2011}
S.~Sachdev,
\newblock \emph{Quantum Phase Transitions},
\newblock Cambridge University Press, 2 edn.,
\newblock \doi{10.1017/CBO9780511973765} (2011).

\bibitem{NicShannon}
N.~Shannon, T.~Momoi and P.~Sindzingre,
\newblock \emph{Nematic order in square lattice frustrated ferromagnets},
\newblock Phys. Rev. Lett. \textbf{96}, 027213 (2006),
\newblock \doi{10.1103/PhysRevLett.96.027213}.

\bibitem{GiampaoloNematic}
S.~M. Giampaolo and B.~C. Hiesmayr,
\newblock \emph{Topological and nematic ordered phases in many-body cluster-ising models},
\newblock Phys. Rev. A \textbf{92}, 012306 (2015),
\newblock \doi{10.1103/PhysRevA.92.012306}.

\bibitem{fradkin_2013}
E.~Fradkin,
\newblock \emph{Field Theories of Condensed Matter Physics},
\newblock Cambridge University Press, 2 edn.,
\newblock \doi{10.1017/CBO9781139015509} (2013).

\bibitem{Witten_TopologicalPhases}
E.~Witten,
\newblock \emph{Fermion path integrals and topological phases},
\newblock Rev. Mod. Phys. \textbf{88}, 035001 (2016),
\newblock \doi{10.1103/RevModPhys.88.035001}.

\bibitem{Giampaolo_2019}
S.~M. Giampaolo, F.~B. Ramos and F.~Franchini,
\newblock \emph{The frustration of being odd: universal area law violation in local systems},
\newblock Journal of Physics Communications \textbf{3}(8), 081001 (2019),
\newblock \doi{10.1088/2399-6528/ab3ab3}.

\bibitem{J_Vannimenus_1977}
J.~Vannimenus and G.~Toulouse,
\newblock \emph{Theory of the frustration effect. ii. ising spins on a square lattice},
\newblock Journal of Physics C: Solid State Physics \textbf{10}(18), L537 (1977),
\newblock \doi{10.1088/0022-3719/10/18/008}.

\bibitem{Chalker2011}
J.~T. Chalker,
\newblock \emph{Geometrically Frustrated Antiferromagnets: Statistical Mechanics and Dynamics}, pp. 3--22,
\newblock Springer Berlin Heidelberg, Berlin, Heidelberg,
\newblock ISBN 978-3-642-10589-0,
\newblock \doi{10.1007/978-3-642-10589-0_1} (2011).

\bibitem{Moessner_Ramirez}
R.~Moessner and A.~P. Ramirez,
\newblock \emph{{Geometrical frustration}},
\newblock Physics Today \textbf{59}(2), 24 (2006),
\newblock \doi{10.1063/1.2186278},
\newblock \eprint{https://pubs.aip.org/physicstoday/article-pdf/59/2/24/16665199/24\_1\_online.pdf}.

\bibitem{QPT_induced_by_TF}
V.~Mari{\'{c}}, S.~M. Giampaolo and F.~Franchini,
\newblock \emph{Quantum phase transition induced by topological frustration},
\newblock Communications Physics \textbf{3}(1) (2020),
\newblock \doi{10.1038/s42005-020-00486-z}.

\bibitem{torre2023longrange}
G.~Torre, J.~Odavić, P.~Fromholz, S.~M. Giampaolo and F.~Franchini,
\newblock \emph{Long-range entanglement and topological excitations} (2023), \eprint{2310.16091}.

\bibitem{Moessner_GeometricalFrustration}
R.~Moessner,
\newblock \emph{Magnets with strong geometric frustration},
\newblock Canadian Journal of Physics \textbf{79}(11-12), 1283 (2001),
\newblock \doi{10.1139/p01-123},
\newblock \eprint{https://doi.org/10.1139/p01-123}.

\bibitem{lacroix}
C.~Lacroix, P.~Mendels and F.~Mila,
\newblock \emph{Introduction to Frustrated Magnetism: Materials, Experiments, Theory},
\newblock ISBN 978-3-642-10588-3,
\newblock \doi{10.1007/978-3-642-10589-0} (2011).

\bibitem{Dong_2016}
J.-J. Dong, P.~Li and Q.-H. Chen,
\newblock \emph{The a-cycle problem for transverse ising ring},
\newblock Journal of Statistical Mechanics: Theory and Experiment \textbf{2016}(11), 113102 (2016),
\newblock \doi{10.1088/1742-5468/2016/11/113102}.

\bibitem{Dong_XY_17}
J.-J. Dong and P.~Li,
\newblock \emph{The a-cycle problem in xy model with ring frustration},
\newblock Modern Physics Letters B \textbf{31}(06), 1750061 (2017),
\newblock \doi{10.1142/S0217984917500610},
\newblock \eprint{https://doi.org/10.1142/S0217984917500610}.

\bibitem{Cabrera86}
G.~G. Cabrera and R.~Jullien,
\newblock \emph{Universality of finite-size scaling: Role of the boundary conditions},
\newblock Phys. Rev. Lett. \textbf{57}, 393 (1986),
\newblock \doi{10.1103/PhysRevLett.57.393}.

\bibitem{Cabrera1987}
G.~G. Cabrera and R.~Jullien,
\newblock \emph{Role of boundary conditions in the finite-size ising model},
\newblock Phys. Rev. B \textbf{35}, 7062 (1987),
\newblock \doi{10.1103/PhysRevB.35.7062}.

\bibitem{Barber87}
M.~N. Barber and M.~E. Cates,
\newblock \emph{Effect of boundary conditions on the finite-size transverse ising model},
\newblock Phys. Rev. B \textbf{36}, 2024 (1987),
\newblock \doi{10.1103/PhysRevB.36.2024}.

\bibitem{Campostrini_Pellissetto_Vicari}
M.~Campostrini, A.~Pelissetto and E.~Vicari,
\newblock \emph{Quantum transitions driven by one-bond defects in quantum ising rings},
\newblock Phys. Rev. E \textbf{91}, 042123 (2015),
\newblock \doi{10.1103/PhysRevE.91.042123}.

\bibitem{Campostrini_2015_Jstat}
M.~Campostrini, A.~Pelissetto and E.~Vicari,
\newblock \emph{Quantum ising chains with boundary fields},
\newblock Journal of Statistical Mechanics: Theory and Experiment \textbf{2015}(11), P11015 (2015),
\newblock \doi{10.1088/1742-5468/2015/11/P11015}.

\bibitem{Dong2018}
J.-J. Dong, Z.-Y. Zheng and P.~Li,
\newblock \emph{Rigorous proof for the nonlocal correlation function in the transverse ising model with ring frustration},
\newblock Phys. Rev. E \textbf{97}, 012133 (2018),
\newblock \doi{10.1103/PhysRevE.97.012133}.

\bibitem{Torre_loschmidt}
G.~Torre, V.~Mari\ifmmode~\acute{c}\else \'{c}\fi{}, D.~Kui\ifmmode~\acute{c}\else \'{c}\fi{}, F.~Franchini and S.~M. Giampaolo,
\newblock \emph{Odd thermodynamic limit for the loschmidt echo},
\newblock Phys. Rev. B \textbf{105}, 184424 (2022),
\newblock \doi{10.1103/PhysRevB.105.184424}.

\bibitem{fate_of_local_order}
V.~Mari\ifmmode~\acute{c}\else \'{c}\fi{}, S.~M. Giampaolo and F.~Franchini,
\newblock \emph{Fate of local order in topologically frustrated spin chains},
\newblock Phys. Rev. B \textbf{105}, 064408 (2022),
\newblock \doi{10.1103/PhysRevB.105.064408}.

\bibitem{Frustration_of_being_odd}
V.~Mari{\'{c} }, S.~M. Giampaolo, D.~Kui{\'{c}} and F.~Franchini,
\newblock \emph{The frustration of being odd: how boundary conditions can destroy local order},
\newblock New Journal of Physics \textbf{22}(8), 083024 (2020),
\newblock \doi{10.1088/1367-2630/aba064}.

\bibitem{Franchini_EffectsOFDefects}
G.~Torre, V.~Mari\ifmmode~\acute{c}\else \'{c}\fi{}, F.~Franchini and S.~M. Giampaolo,
\newblock \emph{Effects of defects in the xy chain with frustrated boundary conditions},
\newblock Phys. Rev. B \textbf{103}, 014429 (2021),
\newblock \doi{10.1103/PhysRevB.103.014429}.

\bibitem{Maric_topological}
V.~Mari{\'{c}}, F.~Franchini, D.~Kui{\'{c}} and S.~M. Giampaolo,
\newblock \emph{Resilience of the topological phases to frustration},
\newblock Scientific Reports \textbf{11}(1) (2021),
\newblock \doi{10.1038/s41598-021-86009-4}.

\bibitem{TF_modify_the_nature_of_QPT}
V.~Marić, G.~Torre, F.~Franchini and S.~M. Giampaolo,
\newblock \emph{{Topological Frustration can modify the nature of a Quantum Phase Transition}},
\newblock SciPost Phys. \textbf{12}, 075 (2022),
\newblock \doi{10.21468/SciPostPhys.12.2.075}.

\bibitem{Catalano22}
A.~G. Catalano, D.~Brtan, F.~Franchini and S.~M. Giampaolo,
\newblock \emph{Simulating continuous symmetry models with discrete ones},
\newblock Phys. Rev. B \textbf{106}, 125145 (2022),
\newblock \doi{10.1103/PhysRevB.106.125145}.

\bibitem{Odavi__2023}
J.~Odavi{\'{c}}, T.~Haug, G.~Torre, A.~Hamma, F.~Franchini and S.~M. Giampaolo,
\newblock \emph{Complexity of frustration: A new source of non-local non-stabilizerness},
\newblock {SciPost} Physics \textbf{15}(4) (2023),
\newblock \doi{10.21468/scipostphys.15.4.131}.

\bibitem{articolo_tesi}
D.~Sacco~Shaikh, M.~Sassetti and N.~Traverso~Ziani,
\newblock \emph{Parity-dependent quantum phase transition in the quantum ising chain in a transverse field},
\newblock Symmetry \textbf{14}(5) (2022),
\newblock \doi{10.3390/sym14050996}.

\bibitem{katsura}
S.~Katsura,
\newblock \emph{Statistical mechanics of the anisotropic linear heisenberg model},
\newblock Phys. Rev. \textbf{127}, 1508 (1962),
\newblock \doi{10.1103/PhysRev.127.1508}.

\bibitem{McCoyXY1}
E.~Barouch, B.~M. McCoy and M.~Dresden,
\newblock \emph{Statistical mechanics of the $\mathrm{XY}$ model. i},
\newblock Phys. Rev. A \textbf{2}, 1075 (1970),
\newblock \doi{10.1103/PhysRevA.2.1075}.

\bibitem{McCoyXY2}
E.~Barouch and B.~M. McCoy,
\newblock \emph{Statistical mechanics of the $xy$ model. ii. spin-correlation functions},
\newblock Phys. Rev. A \textbf{3}, 786 (1971),
\newblock \doi{10.1103/PhysRevA.3.786}.

\bibitem{McCoyXY3}
E.~Barouch and B.~M. McCoy,
\newblock \emph{Statistical mechanics of the $\mathrm{XY}$ model. iii},
\newblock Phys. Rev. A \textbf{3}, 2137 (1971),
\newblock \doi{10.1103/PhysRevA.3.2137}.

\bibitem{LIEB1961407}
E.~Lieb, T.~Schultz and D.~Mattis,
\newblock \emph{Two soluble models of an antiferromagnetic chain},
\newblock Annals of Physics \textbf{16}(3), 407 (1961),
\newblock \doi{https://doi.org/10.1016/0003-4916(61)90115-4}.

\bibitem{NIEMEIJER1967377}
T.~Niemeijer,
\newblock \emph{Some exact calculations on a chain of spins 12},
\newblock Physica \textbf{36}(3), 377 (1967),
\newblock \doi{https://doi.org/10.1016/0031-8914(67)90235-2}.

\bibitem{NIEMEIJER1968313}
T.~Niemeijer,
\newblock \emph{Some exact calculations on a chain of spins 12 ii},
\newblock Physica \textbf{39}(3), 313 (1968),
\newblock \doi{https://doi.org/10.1016/0031-8914(68)90085-2}.

\bibitem{Franchini_2017}
F.~Franchini,
\newblock \emph{An Introduction to Integrable Techniques for One-Dimensional Quantum Systems},
\newblock Springer International Publishing,
\newblock \doi{10.1007/978-3-319-48487-7} (2017).

\bibitem{depasquale}
A.~De~Pasquale and P.~Facchi,
\newblock \emph{$xy$ model on the circle: Diagonalization, spectrum, and forerunners of the quantum phase transition},
\newblock Phys. Rev. A \textbf{80}, 032102 (2009),
\newblock \doi{10.1103/PhysRevA.80.032102}.

\bibitem{Damski_2014}
B.~Damski and M.~M. Rams,
\newblock \emph{Exact results for fidelity susceptibility of the quantum ising model: the interplay between parity, system size, and magnetic field},
\newblock Journal of Physics A: Mathematical and Theoretical \textbf{47}(2), 025303 (2013),
\newblock \doi{10.1088/1751-8113/47/2/025303}.

\bibitem{Tasaki_book}
H.~Tasaki,
\newblock \emph{Physics and Mathematics of Quantum Many-Body Systems},
\newblock ISBN 978-3-030-41264-7,
\newblock \doi{10.1007/978-3-030-41265-4} (2020).

\bibitem{MatthiasVojta_2003}
M.~Vojta,
\newblock \emph{Quantum phase transitions},
\newblock Reports on Progress in Physics \textbf{66}(12), 2069 (2003),
\newblock \doi{10.1088/0034-4885/66/12/R01}.

\bibitem{Anderson1}
P.~W. Anderson,
\newblock \emph{Ground state of a magnetic impurity in a metal},
\newblock Phys. Rev. \textbf{164}, 352 (1967),
\newblock \doi{10.1103/PhysRev.164.352}.

\bibitem{Anderson2}
P.~W. Anderson,
\newblock \emph{Infrared catastrophe in fermi gases with local scattering potentials},
\newblock Phys. Rev. Lett. \textbf{18}, 1049 (1967),
\newblock \doi{10.1103/PhysRevLett.18.1049}.

\bibitem{quartic1}
H.~Sevinçli,
\newblock \emph{Quartic dispersion, strong singularity, magnetic instability, and unique thermoelectric properties in two-dimensional hexagonal lattices of group-va elements},
\newblock Nano Letters \textbf{17}(4), 2589 (2017),
\newblock \doi{10.1021/acs.nanolett.7b00366},
\newblock PMID: 28318269,
\newblock \eprint{https://doi.org/10.1021/acs.nanolett.7b00366}.

\bibitem{quartic2}
C.~M. de~Sterke, A.~F.~J. Runge, D.~D. Hudson and A.~Blanco-Redondo,
\newblock \emph{{Pure-quartic solitons and their generalizations—Theory and experiments}},
\newblock APL Photonics \textbf{6}(9), 091101 (2021),
\newblock \doi{10.1063/5.0059525},
\newblock \eprint{https://pubs.aip.org/aip/app/article-pdf/doi/10.1063/5.0059525/14571872/091101\_1\_online.pdf}.

\bibitem{quartic3}
H.~Triki, A.~Pan and Q.~Zhou,
\newblock \emph{Pure-quartic solitons in presence of weak nonlocality},
\newblock Physics Letters A \textbf{459}, 128608 (2023),
\newblock \doi{https://doi.org/10.1016/j.physleta.2022.128608}.

\bibitem{quartic4}
M.~Olshanii, S.~Choi, V.~Dunjko, A.~Feiguin, H.~Perrin, J.~Ruhl and D.~Aveline,
\newblock \emph{Three-dimensional gross–pitaevskii solitary waves in optical lattices: Stabilization using the artificial quartic kinetic energy induced by lattice shaking},
\newblock Physics Letters A \textbf{380}(1), 177 (2016),
\newblock \doi{https://doi.org/10.1016/j.physleta.2015.09.008}.

\bibitem{as}
M.~Abramowitz and I.~A. Stegun,
\newblock \emph{Handbook of Mathematical Functions with Formulas, Graphs, and Mathematical Tables},
\newblock Dover, New York, ninth dover printing, tenth gpo printing edn. (1964).

\bibitem{Hewitt1979TheGP}
E.~S. Hewitt and R.~E. Hewitt,
\newblock \emph{The gibbs-wilbraham phenomenon: An episode in fourier analysis},
\newblock Archive for History of Exact Sciences \textbf{21}, 129 (1979).

\bibitem{Gottlieb}
D.~Gottlieb and C.-W. Shu,
\newblock \emph{On the gibbs phenomenon and its resolution},
\newblock SIAM Review \textbf{39}(4), 644 (1997),
\newblock \doi{10.1137/S0036144596301390},
\newblock \eprint{https://doi.org/10.1137/S0036144596301390}.

\bibitem{Jordan:1928wi}
P.~Jordan and E.~P. Wigner,
\newblock \emph{{About the Pauli exclusion principle}},
\newblock Z. Phys. \textbf{47}, 631 (1928),
\newblock \doi{10.1007/BF01331938}.

\bibitem{porta}
S.~Porta, F.~Cavaliere, M.~Sassetti and N.~Traverso~Ziani,
\newblock \emph{Topological classification of dynamical quantum phase transitions in the xy chain},
\newblock Scientific Reports \textbf{10}(1), 12766 (2020),
\newblock \doi{10.1038/s41598-020-69621-8}.

\end{thebibliography}

\nolinenumbers

\end{document}